\documentclass[12pt,reqno,a4paper]{amsart}
\usepackage{amssymb,latexsym}
\usepackage{amsthm, amsmath, amsfonts}
\usepackage{hyperref}
\def\be{\begin{equation}}
\def\ee{\end{equation}}
\def\bea{\begin{eqnarray}}
\def\eea{\end{eqnarray}}
\def\bz{\bar z}
\def\bs{\bar s}
\def\p{\partial}
\def\bp{\bar\partial}

\def\bs{\bar s}

\def\bcal_k{\mathcal B_k}
\newcommand{\szego}{Szeg\"o\ }

\newcommand{\kahler}{K\"ahler }

\begin{document}

\title[FQHE on curved backgrounds, free fields and large N]{FQHE on curved backgrounds, free fields and large N}
\author[Frank Ferrari,$^1$ Semyon Klevtsov$^{1,2}$]{Frank Ferrari$^1$ and Semyon Klevtsov$^2$}
\maketitle
{\it\small 
\address{\it$^1$Service de Physique Th\'eorique et Math\'ematique, Universit\'e Libre de Bruxelles}

\address{\it$\phantom{c}$et International Solvay Institutes, Campus de la Plaine, CP 231, 1050 Bruxelles, Belgique}

\address{\it $^2$Mathematisches Institut, Universit\"at zu K\"oln, Weyertal 86-90, 50931 K\"oln, Germany}
}
\vspace{.5cm}

\begin{abstract}
We study the free energy of the Laughlin state on curved backgrounds, starting from the free field representation. A simple argument, based on the computation of the gravitational effective action from the transformation properties of Green functions under the change of the metric, allows to compute the first three terms of the expansion in large magnetic field. The leading and subleading contributions are given by the Aubin-Yau and Mabuchi functionals respectively, whereas the Liouville action appears at next-to-next-to-leading order. We also derive a path integral representation for the remainder terms. They correspond to a large mass expansion for a related interacting scalar field theory and are thus given by local polynomials in curvature invariants. 

\end{abstract}
\tableofcontents
\thispagestyle{empty}

\section{Introduction}

In this paper we study the expansion of the free energy of the Laughlin state on compact Riemann surfaces in a large magnetic field. As is well known, the Laughlin state \cite{L} describes the ground state wave function for the integer and fractional quantum Hall effect. We are specifically interested in the way the free energy depends on the choice of the Riemannian metric. The study of the free energy was initiated in the work of Wiegmann-Zabrodin \cite{ZW,WZ}, who studied large $N$ expansion of the Dyson gas, or $\beta$-ensemble, on the complex plane with an arbitrary magnetic field. They derived the first three terms in the expansion, including the boundary terms, using the method of loop equations. 

Interestingly, the analog of the quantum Hall free energy appears naturally in the context of the Yau-Tian-Donaldson program in \kahler geometry. Namely,  Donaldson \cite{Don2} defined and studied the large $k$ expansion of the determinant of the ${\rm Hilb}_k$-map on any compact \kahler manifold, thus including the case of Riemann surfaces. A similar object was also studied by Berman in \cite{B1,B2}. The Donaldson's expansion follows from the asymptotic expansion of the Bergman kernel \cite{Z,C,Lu,MM} and corresponds to the free-fermion, or $\beta=1$ case. Physically, the Bergman kernel expansion can be understood as an expansion of the density of states on the lowest Landau level (LLL), for the particle in a large magnetic field with flux $k$, on a \kahler manifold. Its physical derivation, using quantum-mechanical path integral methods, can be found in \cite{DK}.

The relation between the expansions of Wiegmann-Zabrodin and Donaldson was recently understood in \cite{K}. The determinant of the ${\rm Hilb}_k$-map of \cite{Don2} corresponds to the partition function of free fermions on LLL on \kahler manifold with arbitrary metric and constant magnetic field. 
This object corresponds to $\beta=1$, or to the case of integer quantum Hall effect.
For any compact Riemann surface, the first five terms of the expansion were computed in \cite{K} (the first two terms were already obtained in \cite{Don2}). Subsequently, the first three terms in the expansion of the free energy of the Laughlin states on the sphere for any $\beta$ were derived in \cite{CLW, CLW1}, generalizing the loop equation method of Ref.\ \cite{ZW} to this case. 

In this paper we derive the free energy expansion for the fractional quantum Hall case by a different and more direct argument. Our method here is not based on Bergman kernel nor on loop equations, but rather on the free field representation of the Laughlin state, and on transformation properties of the Green function under changes of the metric, proved in \cite{FKZ3}. The free field representation of the quantum Hall states is well-known and goes back to the seminal work \cite{MR}. From this point of view, the derivation of the first three terms of the large magnetic flux $k$ expansion amounts to the calculation of the gravitational effective action in a field theory softly breaking conformal invariance, which can be done straightforwardly along the lines of Ref.\ \cite{FKZ3}. We also obtain a new path integral representation of the remainder terms, starting from order $1/k$. 

Now we briefly explain our main result. We consider the Laughlin state on the surface with the constant scalar curvature metric $g_0$, and the same state on the surface with an arbitrary metric $g$, parameterized by the \kahler potential $\phi$ as $g_{z\bz}=g_{0z\bz}+\p_z\bp_{\bz}\phi$. The number of the states on the lowest Landau level is $N_k=k+\chi(M)/2$ on a surface with the Euler characteristic $\chi(M)$. We define the free energy as a logarithm of the ratio of norms of the Laughlin state in the metrics $g$ and $g_0$. Here we quote our main result \eqref{free} for the free energy expansion 
\begin{align}\label{free0}
\mathcal F_\beta[g_0,\phi]=&-2\pi\beta kN_kS_{AY}(g_0,\phi)+\beta\frac k2 S_M(g_0,\phi)-\frac{1-3\beta}{24\pi}S_L(g_0,\phi)+\mathcal R[g_{0},g].
\end{align}
The first three terms here correspond to the Aubin-Yau, Mabuchi and Liouville functionals. The Liouville functional is well-known due to relation to the gravitational anomaly; the first two functionals also appear as gravitational effective actions in two-dimensions \cite{FKZ3}, when conformal invariance is broken. The remainder of the series $\mathcal R[g_{0},g]$ contains the terms of order $1/k$ and less. We derive its path integral representation in sec.\ 5. We also generalize our method to the case when particles have a gravitational spin; in this case the expansion was derived in \cite{CLW1} using loop equation. 

In conformal field theories on curved backgrounds the coefficient in front of the Liouville action is the central charge. The corrections to the free energy \eqref{free0} can also be associated with the various parameters of the quantum Hall system. The coefficient in front of the Aubin-Yau action is the inverse conductance. The Mabuchi term is responsible for the homogeneous part of the anomalous Hall viscosity, see e.g.\ \cite{ASZ,TV,R,RR,HS,W,CLW} and Ref.\ \cite{Ho} for a comprehensive review and complete list of references. The coefficient of Liouville term in this case is conjecturally related to the heat conductance \cite{AG1,AG2}. Let us point out that the effective actions for the quantum Hall system have been also studied recently from the $2+1$ dimensional perspective, see e.g.\ Refs.\ \cite{S,AG3,HKO,GSWW,AG4} for a partial list of references. The scaling limit of Laughlin states in the free field representation on the plane and round sphere and its relation to the Wiegmann-Zabrodin expansion was previously discussed in Ref.\ \cite{DRR}. The geometric response of the quantum Hall effect was first studied in \cite{WZee,FS}. 

Strictly speaking, our calculation here applies to the case of the sphere only, since we study only one Laughlin state, and there are more than one state on the torus \cite{HR} and on the higher-genus surfaces \cite{WN,MR,IL}. The method we develop here can be generalized to Riemann surfaces of any genus, where the new feature is the dependence of the states on the complex structure moduli. We will address this question in future investigation.

The paper is organized as follows. After defining the free field theory on curved backgrounds and computing a relevant correlator of vertex operators in sec.\ 2, we show that it reproduces the Laughlin state on a round sphere in sec.\ 3. In sec.\ 4 we define the free energy and use the transformation properties of the Green function in order to derive the main result, quoted above. In sec.\ 5 we discuss the large $k$ limit of the free energy, and show that the path integral indeed produces the remainder terms starting from order $1/k$. In sec.\ 6 we generalize our method to the case of the conformal spin and in sec.\ 7 we derive some a priori properties of the free energy, which follow from the mathematical formulation of the Laughlin state, using sections of a holomorphic line bundle.

\section{Free field and vertex operators on curved backgrounds}

Consider a compact Riemann surface $M$, equipped with a metric $g=2g_{z\bz}|dz|^2$ with the area normalized as $A=2\pi$. Consider now a free field theory on $(M,g)$ with the action given by the sum of the usual Coulomb-gas term and an additional linear term
\begin{equation}
\label{action}
S(g,\sigma)=\int_M\bigl(2\p_z\sigma\bp_{\bz}\sigma+ib\sigma R\sqrt{g}+2ibk\sigma\sqrt{g}\bigr)d^2z,
\end{equation}
where $b$ is a real number. We use the scalar Gaussian curvature defined by $R=-g_{z\bz}^{-1}\p_z\bp_{\bz}\log\sqrt{g}$ and $\sqrt{g}=2g_{z\bz}$. The relation to the standard Ricci scalar curvature $R^{\rm R}$ is $R=R^{\rm R}/2$. 

The extra linear term in the action\footnote{In \cite{MR} it appears as an insertion in the correlation function, the difference in the normalization of the linear term here and in \cite{MR} is explained by the fact that in the present set-up on a compact surface we fix the total area to $2\pi$.} is proportional to the parameter $k$, which has the dimension ${\rm length}^{-2}$ (magnetic length is defined as $l^2=\hbar/ek$ and we use $e=\hbar=1$ units). Due to the presence of the dimensional parameter one could say that this term ``softly'' breaks the conformal invariance. As we will see now, compared to the pure Coulomb gas case, the only role of this term is in modifying the neutrality condition.

Consider the following non-normalized correlator of some number $N_k$ of vertex operators, inserted at points $z_1,..,z_{N_k}$, which we write using the path integral representation as
\begin{equation}
\label{corr}
Z\bigl(g,\{z_j\}\bigr)=\int e^{ib\sum_{j=1}^{N_k}\sigma(z_j)}e^{-\frac1{4\pi}S(g,\sigma)}\mathcal D_{g}\sigma.
\end{equation}
This correlator is known to be related to the Laughlin states for the fractional quantum Hall effect, this observation goes back to \cite{MR}. The relation between parameters $N_k$ and $k$ is fixed by the usual neutrality condition as follows. We split the field into the zero-mode part and its orthogonal complement
\begin{equation}
\label{decomp}
\sigma=\sigma_0+\tilde\sigma,\quad\int_M\tilde\sigma\sqrt{g}d^2z=0,
\end{equation}
and require that the coefficient in front of the zero mode part in the exponent vanishes: $bN_k-\frac12b\chi(M)-bk=0$. Thus we obtain the following relation between $N_k$ and $k$,
\begin{equation}
N_k=k+\frac{\chi(M)}2,
\end{equation}
where $\chi(M)=\frac1{2\pi}\int_MR\sqrt gd^2z$ is the Euler characteristic of $M$. This being said, we can integrate out the zero mode $\sigma_{0}$ and work with a field $\sigma$ satisfying the constraint $\int\sigma\sqrt{g} d^{2}z = 0$.

The Gaussian path integral \eqref{corr} can be easily computed using standard techniques in free field theory, see e.g.\ \cite{VV,DP}. To this end, we introduce the standard Green function 
\begin{eqnarray}
\label{green}
&&2g_{z\bz}^{-1}\p_z\bp_{\bz} G^{g}(z,y)=-2\pi\delta(z-y)+1,\\\label{intgreen}
&&\int_MG^{g}(z,y)\sqrt{g}d^2y=0,
\end{eqnarray}
and the Green function at coinciding point,
\begin{equation}
\label{reggreen}
G^{g}_R(z,z)=\lim_{z\to y}\bigl(G^{g}(z,y)+\log d_{g}(z,y)\bigr),
\end{equation}
where $d_{g}(z,y)$ is the geodesic distance between the points. Then the path integral \eqref{corr} can be put into standard Gaussian form, shifting $\tilde\sigma$ by 
\begin{eqnarray}
\nonumber
&&\tilde\sigma(z)\to\tilde\sigma(z)+\int_MG^{g}(z,y)j(y)\sqrt{g}d^2y,\\
&&j(y)=2ib\sum_{j=1}^{N_k}\delta(y-z_j)-\frac{ib}{2\pi} R(y),
\end{eqnarray}
and then evaluated. The shift above preserves the orthogonal decomposition \eqref{decomp} due to the property \eqref{intgreen}. The result can be written as 
\begin{eqnarray}
\label{corr1}
\nonumber
&&Z\bigl(g,\{z_j\}\bigr)=\left[\frac{\det'\Delta_g}{2\pi}\right]^{-1/2} \exp\left(-\frac{b^2}{16\pi^2}\iint_MR\sqrt{g}|_zG^{g
}(z,y)R\sqrt{g}|_yd^2z\,d^2y\right)\cdot\\
&& \exp\left(\frac{b^2}{2\pi}\sum_{j=1}^{N_k}\int_MG^{g}(z_j,z)R\sqrt{g}|_zd^2z-b^2\sum_{j\neq m}^{N_k}G^{g}(z_j,z_m)-b^2\sum_{j=1}^{N_k}G^{g}_R(z_j,z_j)\right),
\end{eqnarray}
where $\det'\Delta_g$ is the regularized determinant of the Laplacian in the metric $g$ without including the zero mode.

\section{Laughlin state on the round sphere}

Now, let us consider a case when $M=S^2$ is two-dimensional round sphere with metric of total area $2\pi$
\begin{equation}
g_{0z\bz}=\frac{1}{(1+|z|^2)^2}.
\end{equation}
The scalar curvature equals $R_0=2$ in our conventions. In this case the Green function reads
\begin{equation}
G^{g_0}(z,y)=-\log\frac{|z-y|}{\sqrt{(1+|z|^2)(1+|y|^2)}}-\frac12,
\end{equation}
and the regularized Green function \eqref{reggreen} is just a constant. Equation \eqref{corr1} then yields
\begin{equation}
\label{corr2}
Z\bigl(g_0,\{z_j\}\bigr)=C_0\cdot|\Delta(z)|^{2b^2}\prod_{j=1}^{N_k}(1+|z_j|^2)^{-b^2k},
\end{equation}
where the number of particles $N_k=k+1$ on the sphere, $\Delta(z)=\prod_{i<j}(z_i-z_j)$ is Vandermonde determinant, and $C_0$ is the inverse square root of the regularized determinant of Laplacian, evaluated for the round metric.  After identifying 
\begin{equation}
\label{beta}
\beta=b^2
\end{equation} 
one can recognize in \eqref{corr2} the absolute value squared 
\begin{equation}
\label{Lstate}
|\Psi_L\bigl(g_0,\{z_j\}\bigr)|^2=|\Delta(z)|^{2\beta}\prod_{j=1}^{N_k}(1+|z_j|^2)^{-\beta k}.
\end{equation}
of the Laughlin state \cite{L} for the filling fraction $\nu=1/\beta$, first constructed in the case of the sphere with the round metric and constant magnetic field in Ref.\ \cite{Ho}.

Let us briefly discuss what happens in the higher-genus case. In this case one shall begin with a compactified boson \cite{VV}, and the Laughlin wave functions will have the Green function part and an extra metric-independent factor, depending on the center-of-mass of the system. However, this extra factor will depend on complex structure moduli and on the solenoid phases. In this paper we will be concerned with the dependence on metric only, but our method works in the higher-genus case as well with the appropriate modifications.

\section{Transformation of the metric}

As was already pointed out in \cite{K}, for the Laughlin states on curved backgrounds the \kahler parameterization of the metric is more convenient, than the usual conformal parameterization. For the metrics $g$ and $g_0$ on $M$ their respective \kahler forms $\omega_0=ig_{0z\bz}dz\wedge d\bz$ and $\omega_\phi=ig_{z\bz}dz\wedge d\bz$ differ by a $\p\bp$ of a globally defined scalar function $\phi$, called the \kahler potential,
\begin{align}
\label{kahler}
&\omega_\phi=\omega_0+i\p_z\bp_{\bz}\phi\,dz\wedge d\bz,\\
&\sqrt{g}=\sqrt{g_0}(1+g_{0z\bz}^{-1}\p_z\bp_{\bz}\phi).
\end{align}
Since the metric is everywhere positive on $M$, the \kahler potential must be a subharmonic function, i.e.\ $g_{0z\bz}^{-1}\p_z\bp_{\bz}\phi>-1$. 

We would like to derive the relation between the path integrals $Z\bigl(g_0,\{z_j\}\bigr)$ and $Z\bigl(g,\{z_j\}\bigr)$. This is precisely the problem of computing a gravitational effective action in a theory with a soft breaking of conformal invariance, a problem that we studied in details in \cite{FKZ3}. We can thus straightforwardly follow the strategy used in this reference. We begin with the formula \eqref{corr1} for  $Z\bigl(g,\{z_j\}\bigr)$ and transform it to the new metric, using standard transformation formulas. The determinant of the Laplacian transforms as follows
\begin{equation}
\label{trans1}
\frac{\det'\Delta_g}{\det'\Delta_0}=e^{-\frac1{12\pi}S_L(g_0,g)},
\end{equation}
where $S_L(g_0,g)$ is the Liouville action.
The terms involving the Green function transform as 
\begin{eqnarray}
\label{trans2}&&G^{g}(z,y)-G^{g_0}(z,y)=\frac12(\phi(y)+\phi(z))-2\pi S_{AY}(g_0,\phi),\\
\label{trans3}&&G^g_R(z)-G^{g_0}_R(z)=\frac12\log\frac{\sqrt g}{\sqrt{g_0}}|_z+\phi(z)-2\pi S_{AY}(g_0,\phi),\\
\label{trans4}\nonumber&&\int_MG^g(z_j,z)R\sqrt{g}d^2z-\int_MG^{g_0}(z_j,z)R_0\sqrt{g_0}d^2z=\\&&
=\pi\log\frac{\sqrt g}{\sqrt{g_0}}|_{z_j}+\pi\chi(M)\phi(z_j)-\pi S_M(g_0,\phi)-2\pi^2\chi(M)S_{AY}(g_0,\phi).
\end{eqnarray}
The action functionals here will be defined in a moment. These formulas follow from the definitions \eqref{green} and \eqref{reggreen}, see \cite{FKZ3} for the derivation. The first two formulas here can also be found e.g.\ in \cite{VV,DP}, when restricting to the \kahler parametrization of the metric. Finally, we have the following relation
\begin{align}
\label{trans5}\nonumber\iint_MR\sqrt{g}|_zG^{g}(z,y)R\sqrt{g}|_yd^2z\,d^2y-\iint_MR_0\sqrt{g_0}|_zG^{g_0}(z,y)R_0\sqrt{g_0}|_yd^2z\,d^2y=&\\=2\pi S_L(g_0,\phi)-(2\pi)^2\chi(M)S_M(g_0,\phi),&
\end{align}
which was derived in \cite{FKZ3}. The Aubin-Yau, Mabuchi and Liouville actions have the following form
\begin{align}
S_{AY}(g_0,\phi)=&\,\frac1{(2\pi)^2}\int_M\left(\frac12\phi\,\p_z\bp_{\bz}\phi+\phi\sqrt{g_0}\right)d^2z,\\
S_M(g_0,\phi)=&\,\frac1{2\pi}\int_M\left(\frac{\chi(M)}2\phi\,\p_z\bp_{\bz}\phi\sqrt{g_0}+\phi\bigl(\chi(M)-R_0\bigr)\sqrt{g_0}+\sqrt g\log\frac{\sqrt g}{\sqrt{g_0}}\right)d^2z,\\
\nonumber S_L(g_0,\phi)=&\int_M\left(-\log\frac{\sqrt g}{\sqrt{g_0}}\,\p_z\bp_{\bz}\log\frac{\sqrt g}{\sqrt{g_0}}+R_0\sqrt{g_0}\log\frac{\sqrt g}{\sqrt{g_0}}\right)d^2z=\\
=&\int_M\bigl(-4\eta\,\p_z\bp_{\bz}\eta+2\eta R_0\sqrt{g_0}\bigr)d^2z,
\end{align}
where we also wrote the Liouville action using the conformal parameterization of the metric $\sqrt{g}=e^{2\eta}\sqrt{g_0}$. Since $g$ is uniquely defined by $g_0$ and $\phi$ as in \eqref{kahler}, we use interchangeable notation for metric dependent objects $S(g_0,g):=S(g_0,\phi)$ throughout the paper. The only exception is the Aubin-Yau functional, which is not invariant under the constant shifts of $\phi$, and thus is truly a functional of $\phi$, not $g$. The Mabuchi action was defined in \cite{Mab} and plays a prominent role in \kahler geometry, see e.g.\ \cite{PS} for a review.

Let us now assume that $g_{0}$ is a metric of constant scalar curvature, $R_{0}=\chi(M)$. The above transformation formulas then immediately yield
\begin{equation}
\label{Ztrans}
\log\frac{Z\bigl(g_0,\{z_j\}\bigr)}{Z\bigl(g,\{z_j\}\bigr)}= \beta k\sum_{j=1}^{N_k}\phi(z_j)-2\pi\beta kN_kS_{AY}(g_0,\phi)+\beta\frac k2 S_M(g_0,\phi)+\frac{3\beta-1}{24\pi}S_L(g_0,\phi),
\end{equation}
where $\beta$ is defined in \eqref{beta}.
Let us now use these results to derive our fundamental formula for the Laughlin wave function. For simplicity, we restrict our discussion to the case where $M$ is topologically a sphere. As usual, $g_{0}$ is the round metric and $g$ an arbitrary metric. The absolute value squared of Laughlin wave function, coupled to the metric $g$, has the following form \cite{K,CLW} when expressed in terms of the \kahler potential
\begin{equation}
\label{La}
|\Psi_L(g_0,\phi,\{z_j\})|^2=|\Delta(z)|^{2\beta}\prod_{j=1}^{N_k}(1+|z_j|^2)^{-\beta k}e^{-\beta k\sum_{j=1}^{N_k}\phi(z_j)},
\end{equation}
where in case of the sphere $N_k=k+1$. To compute the norm of this state, we integrate this expression over the positions of the points with the volume form in the metric $g$,
\begin{equation}
\label{normL}
Z_\beta[g_0,\phi]=\int_{M^{N_k}}|\Psi_L(g_0,\phi,\{z_j\})|^2\prod_{j=1}^{N_k}\sqrt g|_{z_j}d^2z_j.
\end{equation}
Now we define the free energy as the logarithm of the ratio, of this norm and the norm of the Laughlin state \eqref{Lstate} for the round metric
\begin{equation}
\label{freedef}
\mathcal F_\beta[g_0,\phi]=\log \frac{Z_\beta[g_0,\phi]}{Z_\beta[g_0,0]}.
\end{equation}
The choice of the sign is unconventional, because this object can also be interpreted as a generating functional for the density correlation functions in FQHE. They can be obtained by taking variations of $\mathcal F_\beta$ with respect to $\phi$. Note that at $\beta=1$ we have $Z_\beta[g_0,0]={\rm const}$, hence $\mathcal F_\beta[g_0,\phi]$ coincides with the one defined in \cite{K}.

Using \eqref{Ztrans} we obtain the following exact formula 
\begin{align}\label{free}
\nonumber
\mathcal F_\beta[g_0,\phi]=&-2\pi\beta kN_kS_{AY}(g_0,\phi)+\beta\frac k2 S_M(g_0,\phi)-\frac{1-3\beta}{24\pi}S_L(g_0,\phi)+\\\nonumber
&+\log\int \left(\int_Me^{i\sqrt\beta\sigma(z)}\sqrt{g}d^2z\right)^{N_k}e^{-\frac1{4\pi}S(g,\sigma)}\mathcal D_g\sigma -\\&-\log\int \left(\int_Me^{i\sqrt\beta\sigma(z)}\sqrt{g_0}d^2z\right)^{N_k}e^{-\frac1{4\pi}S(g_0,\sigma)}\mathcal D_{g_0}\sigma,
\end{align}
By construction the path integral in the second line depends only on the metric $g$ and the path integral in the third line depends only on $g_0$. 

At this point let us take a pause to discuss this result. Note that Eq.\ \eqref{free} holds for any $k$, including finite $k$. If we set $k=0$, the free field action \eqref{action} becomes the standard conformal field theory. In this case the left hand side and the first two terms on the right in \eqref{free} vanish and the rest of the formula reduces to the classical result \cite{P} for the conformal anomaly in the CFT with the central charge $c=1-3b^2$. As we already mentioned, adding the extra linear term in the action \eqref{action} breaks the conformal invariance on the level of the zero modes, modifying the neutrality condition. This adds an extra dimensional parameter $k$ to the problem. In the next section we will take $k$ large and generate the expansion in powers of $1/k$ in Eq.\ \eqref{free}.

\section{Large $k$ limit}

Now we want to study the large $k$ limit in the formula \eqref{free} for the free energy. The first three terms are already organized in the form of the large $k$ expansion, and the result coincides precisely with the result in Ref.\ \cite{CLW}, obtained by the loop equation method. We also have an explicit formula for the remainder terms, of the form
\begin{equation}\label{reamin} \mathcal R[g_{0},g] = \log\mathcal Z[g]-\log\mathcal Z[g_{0}]\, ,
\end{equation}
where we have defined
\begin{equation}\label{fdef} \mathcal Z[g] = \int \left(\int_Me^{i\sqrt\beta\sigma(z)}\sqrt{g}d^2z\right)^{N_k}e^{-\frac1{4\pi}S(g,\sigma)}\mathcal D_g\sigma\, .
\end{equation}
We believe that this explicit path integral representation of the remainder terms could be very useful to actually compute the $1/k$ corrections, but such a detailed analysis is beyond the scope of the present paper and will be presented elsewhere \cite{FK}. Here, we limit ourselves to show that the representation \eqref{fdef} immediately implies that the $1/k$ corrections will be of the expected form, with terms of order $1/k^{p}$ given by an integral of a local polynomial in curvature invariants of dimension $2p+2$.

An elementary way to understand this is to rewrite \eqref{fdef} as
\begin{equation}\mathcal Z[g]=
\int e^{-\frac1{4\pi}S(g,\sigma)+N_k\log\int_Me^{i\sqrt\beta\sigma(z)}\sqrt{g}d^2z}\mathcal D_g\sigma = 
\int e^{-\frac1{4\pi}S_{\rm eff}(g,\sigma)}\mathcal D_g\sigma\, .
\end{equation}
The effective action having a term proportional to $k$ at large $k$, it is natural to consider the expansion around its critical point $\sigma_{c}$, which can be found order by order in $1/k$, 
\begin{equation}
\sigma_c=0+\frac1{2i\sqrt\beta k}(R-\chi)+\frac1{k^2}\left(-\frac1{2(i\sqrt\beta)^3}\Delta R-\frac1{8i\sqrt\beta}\bigl(R^2-\frac1{2\pi}\int_MR^2\sqrt gd^2z\bigr)\right)+O(1/k^{3})\, .
\end{equation}
Here $\Delta$ is one half of the usual Riemannian Laplacian. Now, we expand the effective action around the critical point up to quadratic fluctuations
\begin{multline}
\label{quad}
S_{\rm eff}(g,\sigma_c+\tilde\sigma)=S_{\rm eff}(g,\sigma_c)+
\int_M\left(-\tilde\sigma\Delta\tilde\sigma+\beta k\tilde\sigma^2+\frac\beta 2R\tilde\sigma^2\right)\sqrt gd^2z\\-\frac\beta{8\pi k}\left(\int_M\tilde\sigma R\sqrt gd^2z\right)^2+\mathcal O(\tilde\sigma^3)\, .
\end{multline}
Note that the fluctuating field does not contain zero modes: $\int_M\tilde\sigma\sqrt gd^2z=0$. The value of the effective action at the critical point is
\begin{equation}
S_{\rm eff}(g,\sigma_c)=\frac1{4k}\left(\int_MR^2\sqrt gd^2z-2\pi\chi^2\right)+\mathcal O(1/k^2)\, .
\end{equation}
This contributes to the order $1/k$ term in the free energy. Of course, we also find quantum corrections to the free energy, with contributions from diagrams with arbitrarily many loops at each fixed order in $1/k$. But the crucial point to note on \eqref{quad} is that the quantum fluctuating field $\tilde\sigma$ has a mass squared of order $\beta k$. At large $k$, the contributions from any loop diagram will then reduce to a local integral of a polynomial in curvature invariants, whose dimension is fixed by power counting; at order $1/k^{p}$, we find curvature invariants of dimension $2p+2$. For example, the one-loop diagrams come from the determinant of the quadratic term in \eqref{quad}, whose large mass expansion can be straightforwardly derived from the usual heat kernel expansion and which automatically has the expected form. A similar standard analysis can be performed at any loop order. 

The locality of the large $k$ expansion of $\mathcal Z[g]$ implies that terms of order $k$ must be proportional to the area, which is fixed and thus metric-independent; terms of order $k^{0}=1$ must be proportional to the Euler characteristic, which is also metric-independent; terms of order $1/k$ must be of the form
\begin{equation}
\label{Rsquared}
\frac{c(\beta)}{2\pi k}\int_MR^2\sqrt gd^2z\, ,\end{equation}
for some function $c(\beta)$ of the parameter $\beta$; etc. The form of this general expansion, valid for any $\beta$, is of course consistent with the case $\beta=1$, which was studied in \cite{K}. In particular, the value of $c(1)$ was computed in \cite{K}.

Computing $c(\beta)$, for any $\beta$, is currently under investigation. Note that due to the form of the term \eqref{Rsquared}, we can actually perform the calculation on a round sphere, for which a very explicit formula for $\mathcal Z$ is known. Another interesting remark is to note that $e^{\mathcal Z[g]}$ is the analytic continuation at $s=-N_{k}$ of the holomorphic function
\begin{equation}\label{analyticform} \frac{1}{\Gamma(s)}\int_{0}^{\infty} d t\, t^{s-1}\, 
\int e^{-\frac{1}{4\pi}S(g,\sigma) - t\int_Me^{i\sqrt\beta\sigma(z)}\sqrt{g}d^2z}\mathcal D_g\sigma\, ,
\end{equation}
which is expressed in terms of a standard path integral with the local action functional, reminiscent of the Liouville theory.

\section{Extension to  conformal spin}

Now we extend our derivation to the case when particles have a non-trivial conformal  spin. In this case the absolute value squared of the (unnormalized) Laughlin wave function on the sphere with the metric $g$ has the following form
\begin{equation}
|\Psi_L^s\bigl(g_0,\phi,\{z_j\}\bigr)|^2=|\Delta(z)|^{2\beta}\prod_{j=1}^{\tilde N_k}(1+|z_j|^2)^{-\beta k}\prod_{j=1}^{\tilde N_k}\bigl(\sqrt{g_0}\bigr)^{s}e^{-\beta k\sum_{j=1}^{\tilde N_k}\phi(z_j)+s\sum_{j=1}^{\tilde N_k}\log\frac{\sqrt g}{\sqrt{g_0}}}
\end{equation}
On the mathematical language (see the next section for more details) it means that the Laughlin wave function is now a section of $\tilde N_k$ copies (one for each coordinate $z_j$) of the line bundle $(L^k)^{\otimes \beta}\otimes K_M^{-s}$, where $K_M^{-1}$ is anticanonical line bundle\footnote{this is the sign choice for the sphere; for genus $g>1$ the sign of $s$ is usually switched and canonical line bundle is used instead.}. The number of particles $\tilde N_k$, or equivalently, the dimension of the space of holomorphic sections $H^0(M,L^k\otimes K_M^{-s/\beta})$ in this case (see e.g. \cite{MM}) is given by
\begin{equation}
\tilde N_k=k+\frac{\chi(M)}{2\beta}(\beta+2s).
\end{equation}

To treat this generalization, we use the following modified version of the free field action \eqref{action},
\begin{equation}
S_q(g,\sigma)=\int_M\bigl(- 2\sigma\p_z\bp_{\bz}\sigma+iq\sigma R\sqrt{g}+2ib\lambda\sigma\sqrt{g}\bigr)d^2z,
\end{equation}
where the appropriate choice of parameters $b,q$ is 
\begin{equation}
b^2=\beta,\quad q=\frac{\beta+2s}{\sqrt \beta}.
\end{equation}
As before, the value of $\lambda$ is fixed to be
\begin{equation}
\lambda=\tilde N_k-\frac{\chi(M)}2\frac qb=k
\end{equation}
so that the neutrality condition holds for the correlation function
\begin{equation}
Z_s\bigl(g,\{z_j\}\bigr)=\int e^{ib\sum_{j=1}^{\tilde N_k}\sigma(z_j)}e^{-\frac1{4\pi}S(g,\sigma)}\mathcal D_{g}\sigma.
\end{equation}
This path integral can be computed as before,
\begin{eqnarray}
\label{corrq}
\nonumber
&&Z_s\bigl(g,\{z_j\}\bigr)=\left[\frac{\det'\Delta_g}{2\pi}\right]^{-1/2} \exp\left(-\frac{q^2}{16\pi^2}\iint_MR\sqrt{g}|_zG^{g
}(z,y)R\sqrt{g}|_yd^2z\,d^2y\right)\cdot\\
&& \exp\left(\frac{bq}{2\pi}\sum_{j=1}^{N_k}\int_MG^{g}(z_j,z)R\sqrt{g}|_zd^2z-b^2\sum_{j\neq m}^{N_k}G^{g}(z_j,z_m)-b^2\sum_{j=1}^{N_k}G^{g}_R(z_j,z_j)\right),
\end{eqnarray}
to be compared with \eqref{corr1}.
The transformation property \eqref{Ztrans} now reads
\begin{align}\nonumber
\label{Ztrans1}
Z_s\bigl(g_0,\{z_j\}\bigr)=Z_s\bigl(g,\{z_j\}\bigr)\,&\cdot\, e^{\beta k\sum_{j=1}^{\tilde N_k}\phi(z_j)- s\sum_{j=1}^{\tilde N_k}\log\frac{\sqrt g}{\sqrt{g_0}}|_{z_j}}\\&\cdot e^{-2\pi\beta k\tilde N_kS_{AY}(g_0,\phi)+(\beta+2s)\frac k2 S_M(g_0,\phi)+\frac{3(\beta+2s)^2-\beta}{24\pi\beta}S_L(g_0,\phi)}.
\end{align}
Denoting the norm of the Laughlin wave function with the spin as 
\begin{equation}
Z_{\beta,s}[g_0,\phi]=\int_{M^{\tilde N_k}}|\Psi^s_L(g_0,\phi,\{z_j\})|^2\prod_{j=1}^{\tilde N_k}\sqrt g|_{z_j}d^2z_j.
\end{equation}
we obtain the generalization of Eq.\ \eqref{free},
\begin{align}\label{free1}
\nonumber
\mathcal F_{\beta,s}[g_0,\phi]=&-2\pi\beta k\tilde N_kS_{AY}(g_0,\phi)+ \bigl(\beta+2s\bigr)\frac k2 S_M(g_0,\phi)-\frac{\beta-3(\beta+2s)^2}{24\pi\beta}S_L(g_0,\phi)\,+\\\nonumber
&+\log\int \left(\int_Me^{i\sqrt\beta\sigma(z)}\sqrt{g}d^2z\right)^{\tilde N_k}e^{-\frac1{4\pi}S_q(g,\sigma)}\mathcal D_g\sigma-\\&
-\log\int \left(\int_Me^{i\sqrt\beta\sigma(z)}\sqrt{g_0}d^2z\right)^{\tilde N_k}e^{-\frac1{4\pi}S_q(g_0,\sigma)}\mathcal D_{g_0}\sigma,
\end{align}
The first three terms here coincide with the loop equation result of Ref.\ \cite{CLW1}.

\section{General form of the free energy}

In previous sections we computed the first three terms of the expansion of the free energy and gave a path integral representation for the remainder terms of the expansion. Our arguments there depend on the choice of the background metric $g_0$, which we have chosen to be the constant scalar curvature metric. In this section we would like to show that the result actually holds for an arbitrary choice of the background metric $g_0$. As we will see, this will follow from certain cocycle conditions, satisfied by the free energy.

So we do not assume any condition on the background metric, and $g_0$ and $g$ will be two arbitrary metrics in the same \kahler class. Also, the arguments in this section apply not only to the Riemann surface case, but to higher-dimensional compact \kahler manifolds of complex dimension $n$, where the lowest Landau level (LLL) wave functions \cite{DK} and Quantum Hall partition function \cite{K} can also be defined. Quantum Hall effect in higher-dimensions was also considered in \cite{KN1}. 

Recall that the single particle LLL wave functions on $(M,g)$ are associated with the sections $s_j(z)$ of the positive holomorphic line bundle $L^k$ (the argument below is also valid for the tensor product with the canonical line bundle $K_M^{-s}$). For a choice of the background metric $g_0$, the magnetic field is given by the $(1,1)$-form $kg_{0z\bz}=-\p_z\bp_{\bz}\log h_0^k$, where $h_0^k$ is the Hermitian metric on $L^k$. Consider an orthonormal basis of the wave functions with respect to the background metric
\begin{equation}
\frac1{2\pi}\int_{M}\bs_i(\bz)s_j(z)h_0^k(z,\bz)\,\sqrt{g_0}d^2z=\delta_{ij}.
\end{equation}
Following \cite{Don2,B1,B2,K} consider the following partition function
\begin{equation}
\label{part}
Z_\beta[g_0,g]:=Z_\beta[g_0,\phi]=\int_{M^{N_k}}|\det s_i(z_j)|^{2\beta}\prod_{j=1}^{N_k}h_0^{\beta k}(z_j)e^{-\beta k\sum_j\phi(z_j)}\prod\sqrt g|_{z_j}d^2z_j.
\end{equation}
In particular on $S^2$ we have $s_j(z)=\sqrt{N_kC_k^{j-1}}z^{j-1}$ and $j=1,...,N_k=k+1$. Thus the partition function above coincides with the norm of the Laughlin wave function \eqref{normL} for $M=S^2$, up to an inessential numerical constant. At $\beta=1$ the integral above admits a determinantal representation \cite{Don2}, and therefore satisfies the cocycle condition, i.e.\ for any three metrics $g_0,g$ and $g_1$ in the same \kahler class (in two dimensions, of the same area) we have \cite{K},
\begin{equation}
\label{cocycle}
Z_1[g_0,g]Z_1[g,g_1]=Z_1[g_0,g_1].
\end{equation}
Now we show that the following combination
\begin{equation}
\label{BI}
\frac{Z_\beta[g_0,g]}{\bigl(Z_1[g_0,g]\bigr)^\beta}
\end{equation}
is in fact independent of $g_0$, i.e.\ background independent. To this end, let us construct another basis of sections,  $t_j(z)$, orthonormal with respect to the metric $g_{z\bz}=-\frac1k\p_z\bp_{\bz}\log h^k$, where $h^k=h_0^ke^{-k\phi}$ is the corresponding Hermitian metric on the line bundle $L^k$,
\begin{equation}
\frac1{2\pi}\int_{M}\bar t_i(\bz)t_j(z)h^k(z,\bz)\,\sqrt{g}d^2z=\delta_{ij}.
\end{equation}
There exists a linear transformation between the bases
\begin{equation}
s_i=A_{ij}t_j.
\end{equation}
From the definition \eqref{part}, it follows that
\begin{equation}
\frac{Z_\beta[g_0,g]}{\bigl(Z_1[g_0,g]\bigr)^\beta}=\frac{(\det A^\dagger A)^\beta Z_\beta[g,g]}{(\det A^\dagger A)^\beta\bigl(Z_1[g,g]\bigr)^\beta}=Z_\beta[g,g],
\end{equation}
and the latter is, by construction, independent of the background metric $g_0$. Let us stress, that the background independent partition function $Z_\beta[g,g]$ has the following natural meaning. It is constructed starting from the one-particle states, which are normalized with respect to the metric $g$, and not $g_0$ as $Z_\beta[g_0,g]$. From the expansion \eqref{free} it follows that the background independent partition function satisfies the following transformation formula
\begin{equation} 
\log \frac{Z_\beta[g,g]}{Z_\beta[g_0,g_0]}= -\frac{1-\beta}{24\pi}S_L(g_0,g)+\mathcal O(1/k).
\end{equation}
At infinite $k$ this formula becomes exact and coincides with the confromal anomaly of a CFT with the central charge $c=\beta-1$.

Now we show that the normalized partition function constructed in \eqref{freedef},
\begin{equation}
\tilde Z_{\beta}[g_0,g]=\frac{Z_\beta[g_0,g]}{Z_\beta[g_0,g_0]}\, ,
\end{equation}
also satisfies the cocycle condition
\begin{equation}
\label{cocycle1}
\tilde Z_\beta[g_0,g]\tilde Z_\beta[g,g_1]=\tilde Z_\beta[g_0,g_1]\, .
\end{equation}
Using the background independence of \eqref{BI}, and the cocycle condition at $\beta=1$ \eqref{cocycle}, we get
\begin{align}\nonumber
\tilde Z_\beta[g_0,g]\tilde Z_\beta[g,g_1]&=\frac{Z_\beta[g_0,g]}{Z_\beta[g_0,g_0]}\frac{Z_\beta[g,g_1]}{Z_\beta[g,g]}=
\frac{Z_\beta[g_0,g]}{Z_\beta[g_0,g_0]}\frac{Z_\beta[g,g_1]}{Z_\beta[g_0,g]}(Z_1[g_0,g])^\beta=\\&\nonumber=\frac{Z_\beta[g,g_1]}{Z_\beta[g_0,g_0]}(Z_1[g_0,g])^\beta=\frac1{Z_\beta[g_0,g_0]}\frac{Z_\beta[g,g_1]}{(Z_1[g,g_1])^\beta}(Z_1[g,g_1]Z_1[g_0,g])^\beta=\\\nonumber&=\frac1{Z_\beta[g_0,g_0]}\frac{Z_\beta[g_0,g_1]}{(Z_1[g_0,g_1])^\beta}(Z_1[g,g_1]Z_1[g_0,g])^\beta=\frac{Z_\beta[g_0,g_1]}{Z_\beta[g_0,g_0]}=\tilde Z_\beta[g_0,g_1].
\end{align}
This generic result has two important consequences, First, if the expansion of the modified free energy exists, it has the form
\begin{equation}
\log \tilde Z_\beta[g_0,g]=\sum_{m=0}^\infty k^{2-m}S_m(g_0,g),
\end{equation}
where the functionals $S_m(g_0,g)$ satisfy additive cocycle identity 
\begin{equation}
\label{cocycleadd}
S_m(g_0,g_1)=S_m(g_0,g)+S_m(g,g_1).
\end{equation}
This is known to be true for the Aubin-Yau, Mabuchi and Liouville actions, see e.g. \cite{K} and references therein. It also holds trivially for the corrections of the type \eqref{Rsquared}, which are differences of the local density of curvature invariants. 

Second, the expansion \eqref{free}, that we obtained under the assumption that $g_0$ is the round metric on the sphere, is valid verbatim if $g_0$ is an arbitrary metric of the same area. This follows directly form \eqref{cocycleadd}.

\section{Discussion}

In this paper we develop a new method to derive large magnetic flux expansion of the norm of Laughlin state, based on the free field representation and on the transformation properties of the correlation functions, derived in \cite{FKZ3}. We derived the first three terms in the expansion, which agree with the loop equation result \cite{CLW}. We also propose a new representation of the remainder terms in the expansion, as a path integral in certain interacting field theory.

Much remains to be done. Our observation opens several intriguing possibilities. The method here can be applied in the higher-genus case. We expect the pure metric-dependent terms in the expansion to be the same, but deriving the moduli-dependent terms will be of particular interest. The scaling limits of other quantum Hall states, such as the Pfaffian state \cite{MR}, could potentially be tackled by our method. Various correlation functions in quantum Hall admit free field representation and thus could be within the reach of our approach.

\vspace{.5cm}

{\bf Acknowledgments} We would like to thank E.~Bettelheim, R.~Bondesan, T.~Can, J.~Dubail, I.~Gruzberg, M.~Laskin and  P.~Wiegmann for useful discussions and comments. FF is supported in part by the belgian FRFC (grant 2.4655.07) and IISN (grant 4.4511.06 and 4.4514.08). SK is supported by the postdoctoral fellowship from the Alexander von Humboldt Foundation. He is also supported in part by the grants RFBR 12-01-00482, NSh-1500.2014.2.

\end{document}